\documentclass[runningheads]{llncs}
\usepackage[T1]{fontenc}
\usepackage{graphicx}
\usepackage{amsmath}
\usepackage{booktabs}
\usepackage{appendix}
\usepackage{float}
\usepackage{subfig}
\usepackage[misc]{ifsym}
\usepackage{makecell}
\usepackage{booktabs}
\usepackage[colorlinks,
            linkcolor=red,
            anchorcolor=blue,
            citecolor=blue
            ]{hyperref}
\begin{document}
\begin{sloppypar}
\title{BM-PAW: A Profitable Mining Attack in the PoW-based Blockchain System}
%
%

\author{Junjie Hu \and
Na Ruan $^{(\textrm{\Letter})}$}

\authorrunning{Hu et al.}


\institute{Department of Computer Science and Engineering, Shanghai Jiao Tong University,
\\ Shanghai 200240, China\\
\email{\{nakamoto, naruan\}@sjtu.edu.cn}}

\maketitle              
\begin{abstract}
Mining attacks enable an adversary to procure a disproportionately large portion of mining rewards by deviating from honest mining practices within the PoW-based blockchain system. In this paper, we demonstrate that the security vulnerabilities of PoW-based blockchain extend beyond what these mining attacks initially reveal. We introduce a novel mining strategy, named BM-PAW, which yields superior rewards for both the attacker and the targeted pool compared to the state-of-the-art mining attack, PAW. BM-PAW attackers are incentivized to offer appropriate bribe money to other targets, as they comply with the attacker's directives upon receiving payment. We further find the BM-PAW attacker can circumvent the “miner's dilemma” through equilibrium analysis in a two-pool BM-PAW game scenario, wherein the outcome is determined by the attacker's mining power. We finally propose practical countermeasures to mitigate these novel pool attacks.

\end{abstract}
\section{Introduction}
Bitcoin~\cite{nakamoto2008bitcoin}, a decentralized cryptocurrency grounded in blockchain technology, was introduced by Satoshi Nakamoto in November 2008. Distinct from conventional currencies, Bitcoin does not depend on designated currency institutions for issuance. Instead, it is produced through extensive computations in accordance with predefined algorithms. The PoW-based blockchain system employs a distributed database encompassing numerous nodes across the entire peer-to-peer network to verify and record all transactions. It also adopts cryptographic designs to safeguard the security of the entire currency circulation process. The decentralization of the peer-to-peer network and consensus algorithms ensures that the currency value cannot be artificially manipulated through the mass creation of PoW-based blockchain. The cryptographic-based designs allow PoW-based tokens to be transferred or paid for solely by its rightful owners.

In the PoW-based blockchain system, participants known as miners can obtain rewards by appending transaction records to the ledger, referred to as the blockchain. This process necessitates miners solving cryptographic puzzles, serving as a proof of work~\cite{coin}. The miner who first solves the puzzle and generates a valid block is eligible to receive block rewards, which amounted to 3.125 Bitcoins in November 2024. This activity of solving cryptographic puzzles and generating blocks is termed the “mining process”. Occasionally, due to network communication delays, two or more blocks may be generated and published simultaneously, leading to a fork in the blockchain. To ensure consistency, the system selects one of these branches to ultimately become the main chain. Upon recognizing the longest chain, miners on other branches redirect their attention and mining power to the main chain. In the Bitcoin system, the difficulty of solving cryptographic puzzles is adjusted every two weeks to maintain a constant average block generation time of approximately 10 minutes. However, given the current mining power's hash rate exceeding $3.3 \times 10^{20}$ Hash/s~\cite{Bitinfocharts}, it may take an individual miner several months or even years to solve a mining puzzle~\cite{schrijvers2017incentive}. Consequently, to achieve stable income, miners tend to unite and form mining pools. Most mining pools have a designated pool manager who is responsible for allocating tasks and distributing rewards. When a mining pool discovers a block, the reward is shared among the pool's miners based on their contributions, specifically the number of shares they have submitted.

Cryptocurrencies, possessing intrinsic monetary value, inherently become a lucrative target for cyberattacks. Despite the robust security design of PoW-based blockchain, prior research has demonstrated that it may be more financially beneficial for an attacker to deviate from honest mining strategies. Such deviations include adopting higher-level mining tactics associated with selfish mining~\cite{eyal2018majority,nayak2016stubborn,li2021semi,li2022semi}, engaging in block withholding (BWH)~\cite{courtois2014subversive}, executing fork after withholding (FAW)~\cite{kwon2017selfish}, and employing power adjusting withholding (PAW)~\cite{gao2019power}.

In this paper, we introduce a novel mining attack targeting pool mining, called Power Adjusting Withholding Attack with Bribe Money (BM-PAW), which promises the attacker greater profits than those attainable through PAW attack. Our underlying insight is that it is beneficial for the attacker to offer an appropriate bribe to other miners, as these miners may comply with the attacker's directives if they accept the payment. We term such payment “bribe money” (BM) and refer to the induced miners as the “target pool”. It is important to note that BM differs from bribery in the context of bribery selfish mining (BSM). For example, the attacker in our scenario has the authority to instruct the target pool to discard the Full Proof of Work (FPoW) or mine on the attacker's branch. BM-PAW extends the traditional PAW attack by incorporating a unique mining strategy centered on BM. The rationale behind BM-PAW is that the attacker consistently achieves additional profits when they possess greater mining power or their competitor has lesser mining power. In BM-PAW, the attacker initially infiltrates the target pool to persuade it to accept the bribe and participate in the BM-PAW attack. During this phase, the attacker negotiates with the target pool to determine the proportion of the reward that will be offered as BM per block. Once the target accepts the bribe, it must adhere to the attacker's instructions, either mining on the attacker's FPoW or discarding blocks in different forking competitions. We demonstrate that with the appropriate selection of BM, it is feasible to achieve incentive compatibility between the attacker and the target pool. Specifically, both the target and the attacker can obtain higher rewards than those attainable through PAW.

In this paper, we propose and discuss the following vital questions:

\begin{itemize}
    \item[$\bullet$] \textit{First, how to adjust the attacker’s infiltration mining power to maximize his rewards?}
    \item[$\bullet$] \textit{Moreover, how to choose the proper bribe money to achieve incentive compatibility between the attacker and the target pool?}
    \item[$\bullet$] \textit{Finally, what equilibrium will the BM-PAW game of different types of participants eventually reach?}
\end{itemize}

\subsection{Our Contributions}

In our research, we introduce a novel mining attack denominated as BM-PAW. In this attack scenario, the attacker is incentivized to offer an appropriate amount of bribe money to a target pool, which in turn discards forks or mines on the attacker's FPoW. To address the initial inquiry, we derive the expected reward for the attacker and formulate the optimal infiltration mining power adjustment by solving the corresponding optimization equation. Our quantitative analysis and simulations reveal that both the attacker and the target pool can achieve higher rewards compared to the advanced and profitable mining attack known as PAW. This finding underscores the vulnerability of pooled mining in the PoW-based blockchain network.

Subsequently, we undertake a formal analysis of the BM-PAW attack across various scenarios. Our analysis specifically focuses on both a singular victim pool and a game involving multiple BM-PAW attackers. To address the second question, we individually scrutinize the rewards accrued by both the attacker and the target pool. We incorporate the constraints that the minimum bribe amount must suffice to ensure the target pool's reward in BM-PAW exceeds that in PAW, while the maximum bribe amount must guarantee that the attacker's reward in BM-PAW is greater than in PAW. Furthermore, we demonstrate that with an apt selection of bribe money, it is feasible to attain incentive compatibility between the attacker and the target pool, fostering a mutually beneficial outcome. Concurrently, simulations corroborate that with an appropriate bribe amount, both the BM-PAW attacker and the target pool can garner additional rewards surpassing those obtained in PAW.

Furthermore, we elucidate the possibility of two mining pools employing the BM-PAW attack against one another. To address the third question, we delve into the analysis of the Nash equilibrium in a game involving two pools engaging in BM-PAW. Our findings indicate that the BM-PAW attacker can circumvent the “miner's dilemma”, where the outcome is traditionally dictated by the pool size, with the larger pool emerging victorious. Notably, under the Nash equilibrium, one pool can meet the criteria for victory and achieve higher profits compared to honest mining, thereby overcoming the “miner's dilemma”.

Finally, we deliberate upon practical countermeasures to mitigate the impact of advanced attacks involving bribe money, thereby fostering innovative ideas for future research directions.

\subsection{Related Work}

\noindent\textbf{BWH Attacks.} The attacker can adopt the BWH attack to destroy reward for the victim pool~\cite{courtois2014subversive}. He divides the mining power into innocent mining pool and infiltration mining pool. When the infiltration pool finds a valid block (full proof of work, FPoW), he withholds it and continues to submit other shares (partial proof of work, PPoW) to obtain the share reward. The victim mining pool will never get reward from the attacker's infiltration mining. Hence, the victim pool will suffer a loss. Other miners, including the attacker’s innocent pool, will gain more reward for the loss of the victim pool.~\cite{luu2015power} indicates that when the attacker splits his mining power properly, BWH attack is more profitable than selfish mining. However, when multiple independent pools adopt BWH attack against each other, they will encounter the “miner's dilemma”~\cite{eyal2015miner}.

\noindent\textbf{FAW Attacks.} The FAW attack extends BWH with forks~\cite{kwon2017selfish}. In brief, the BWH attacker will discard the discovered FPoW, while the FAW attacker will reserve the FPoW. When other miners (not in the victim pool) find a valid blocks, the attacker will release and submit the previously reserved FPoW and generate a fork. An FAW attacker can get more reward than in BWH and avoid miner's dilemma. When the attacker's FPoW is not selected as the main chain, FAW degenerates into BWH.~\cite{gao2019power} combines mining power adjustment strategy with FAW attack, allowing the attacker to adjust mining power dynamically between innocent mining and infiltration mining. Furthermore,~\cite{yang2022if} proposes FWAP where the mining pool pays the attacker for withholding a fork, which brings higher reward for both the attacker and the paying pool.

\noindent\textbf{Bribery Attacks.} The bribery attack can increase the probability of the attacker's branch being selected as the main chain in forking competition~\cite{bonneau2016buy}, which is always considered to combine with other attacks with forks~\cite{karame2012double,zhang2022insightful}. Bribery attack can be launched in a less visible way~\cite{mccorry2019smart}.~\cite{gao2019power} combines bribery attack with selfish mining, which indicates that BSM could bring $10\%$ extra rewards to the attacker than honest mining. However, BSM may cause the “venal miner's dilemma”.~\cite{yang2020ipbsm} proposes an optimal BSM to avoid the “venal miner's dilemma”, where the miner is considered perfectly rational.~\cite{wang2020optimal} proposes a mixed scenario where the attacker alternates his strategies between BWH, FAW, and PAW. The mixed strategy is proved to be much higher in revenue than honest mining.

\section{Preliminaries}
\subsection{Mining Process}
The issuance process of Bitcoin is facilitated by the Bitcoin system, which generates a predetermined quantity of Bitcoins as a reward for miners, thereby assigning the role of currency issuer to the miners. This process of creating a new block is commonly referred to as mining. All Bitcoin transactions must be bundled into blocks and subsequently recorded in the blockchain ledger. The miner who first identifies a nonce that satisfies the specified difficulty criteria is eligible to receive the coinbase reward (also known as the block reward). The mining process serves as an incentive for miners to uphold the security of the blockchain. The total supply of Bitcoins is capped at 21 million. Initially, each miner who successfully publishes a block receives a coinbase reward of 50 Bitcoins, which undergoes a halving event every four years. It is anticipated that by 2104, the coinbase reward will become too small to be further subdivided, marking the completion of the issuance of all Bitcoins.
\subsection{Forks}
When multiple valid blocks are concurrently generated, a forking event occurs. An honest miner will designate the first received valid block as the header~\cite{decker2013information}. Subsequently, one of the branches will emerge victorious in the competition and ultimately become the main chain. A miner who successfully publishes a block on the main chain will be awarded the corresponding coinbase reward, whereas other miners involved in the forked branches will receive no reward. It is worth noting that forks can also be intentionally induced, such as through selfish mining or FAW attacks.
\subsection{Mining Pool}
As the investment in mining power for PoW-based blockchain continues to escalate, the likelihood of an individual miner discovering a valid block diminishes significantly. Consequently, miners increasingly tend to join organizations known as mining pools. Typically, a mining pool comprises a pool manager and numerous peer miners who collaborate to solve the same cryptographic puzzle. Upon successfully generating a valid block, the mining pool's participants will divide the reward according to a predefined distribution protocol, such as Pay Per Share (PPS), Pay Per Last N Shares (PPLNS), or Pay Proportionally (PROP)~\cite{zolotavkin2017incentive}, among others. In theoretical analysis, a miner's reward is directly proportional to their mining power. Therefore, by joining a mining pool, a miner can significantly mitigate the variability in profit. Currently, the majority of blocks in Bitcoin are generated by mining pools, including AntPool~\cite{Antpool}, Poolin~\cite{Poolin}, and F2Pool~\cite{gencer2018decentralization}.

\section{Threat Model and Assumption}
\subsection{Threat Model}
An attacker may manifest as an individual miner or as a mining pool comprising multiple miners. Honest miners, motivated by profit maximization, are inclined to adopt optimal mining strategies to augment their earnings without initiating any mining attacks. Furthermore, the attacker has the capability to establish multiple identities through Sybil attacks and subsequently participate in various open mining pools utilizing distinct accounts and identifiers. Concurrently, the attacker's mining power is constrained to prevent the execution of a 51\% attack. The attacker can allocate their mining power across different strategies, including honest mining within an innocent mining pool (behaving as an honest miner), infiltration mining within other pools, or engaging in other mining attack methodologies. More precisely, within our proposed BM-PAW model, the attacker allocates their mining power to both innocent mining pools and infiltration mining pools. Ultimately, the attacker can create Sybil nodes within the network to prioritize the dissemination of their generated blocks, thereby enhancing the likelihood of their branch being selected as the main chain in the event of a fork.
\subsection{Assumptions}
To facilitate our analysis, we adopt a series of plausible assumptions that align with those employed in the study of other mining attacks, such as selfish mining~\cite{eyal2018majority}, greedy mining~\cite{hu2024greedy}, and bribery stubborn mining~\cite{hu2023novel}.

\begin{itemize}
    \item[$\bullet$] Normalization of Mining Power: We normalize the total mining power of the system to unity. Within this framework, the attacker's normalized mining power is bounded between 0 and 0.5, excluding both endpoints. This constraint is intended to preclude the possibility of a 51\% attack.
    \item[$\bullet$] Profit-Driven Miner Behavior: Miners are motivated by profit maximization. An honest miner may adopt what they perceive as the optimal mining strategy to enhance their earnings but refrain from initiating mining attacks. This assumption is grounded in the recognition that miners are honest but inherently profit-driven. In scenarios where the blockchain forks, and the lengths of the competing branches are equivalent, miners have the discretion to select any branch.
    \item[$\bullet$] Absence of Unintentional Forks: We assume that unintentional forks are absent in the PoW-based blockchain system. This presumption is justified by the negligible probability of such events occurring, estimated to be approximately 0.41\%~\cite{gervais2016security}. In conjunction with Assumption 1, the expected reward for a miner is equivalent to their probability of discovering a valid block in each round. Given the exponential distribution of block discovery times~\cite{eyal2016bitcoin}, where the mean is inversely proportional to a miner's mining power, the probability of a miner finding a valid block corresponds to their normalized mining power.
    \item[$\bullet$] Normalization of Block Reward: We normalize the block reward (coin-base reward) to unity, rather than the current Bitcoin reward of 3.125 BTCs. In our analysis, miners' rewards are both expected and normalized.
    \item[$\bullet$] Reward Distribution in Mining Pools: Upon discovering an FPoW, a pool manager will propagate the valid block to obtain the block reward and subsequently distribute it among pool miners based on the number of submitted shares (PPoW and FPoW) in each round.
    \item[$\bullet$] Bribery in Forking Competition: Certain miners may offer bribe money to the target to increase the likelihood of their FPoW being selected as the main chain in a forking competition. This assumption acknowledges the potential for strategic bribery in the context of blockchain forking.
\end{itemize}

\section{One-pool: BM-PAW}
\subsection{Overview}
We introduce a novel mining attack strategy that leverages bribe money as a strategic mechanism. Specifically, the attacker offers bribe money to target miners in order to enhance the probability of their FPoW being selected as the main chain in a forking competition. By integrating this bribe money strategy with PAW attack, we derive a unique mining strategy termed BM-PAW. The fundamental principle underlying our innovative BM-PAW attacks is the utilization of monetary incentives to target miners for the purpose of maximizing profit. The attacker possesses the capability to dynamically allocate their mining power between honest mining and infiltration mining. Furthermore, in scenarios involving forking competitions, the attacker can employ additional monetary incentives to target miners.
\subsection{Method}
\begin{figure}[t]
  \centering
  \includegraphics[width=1.0\linewidth]{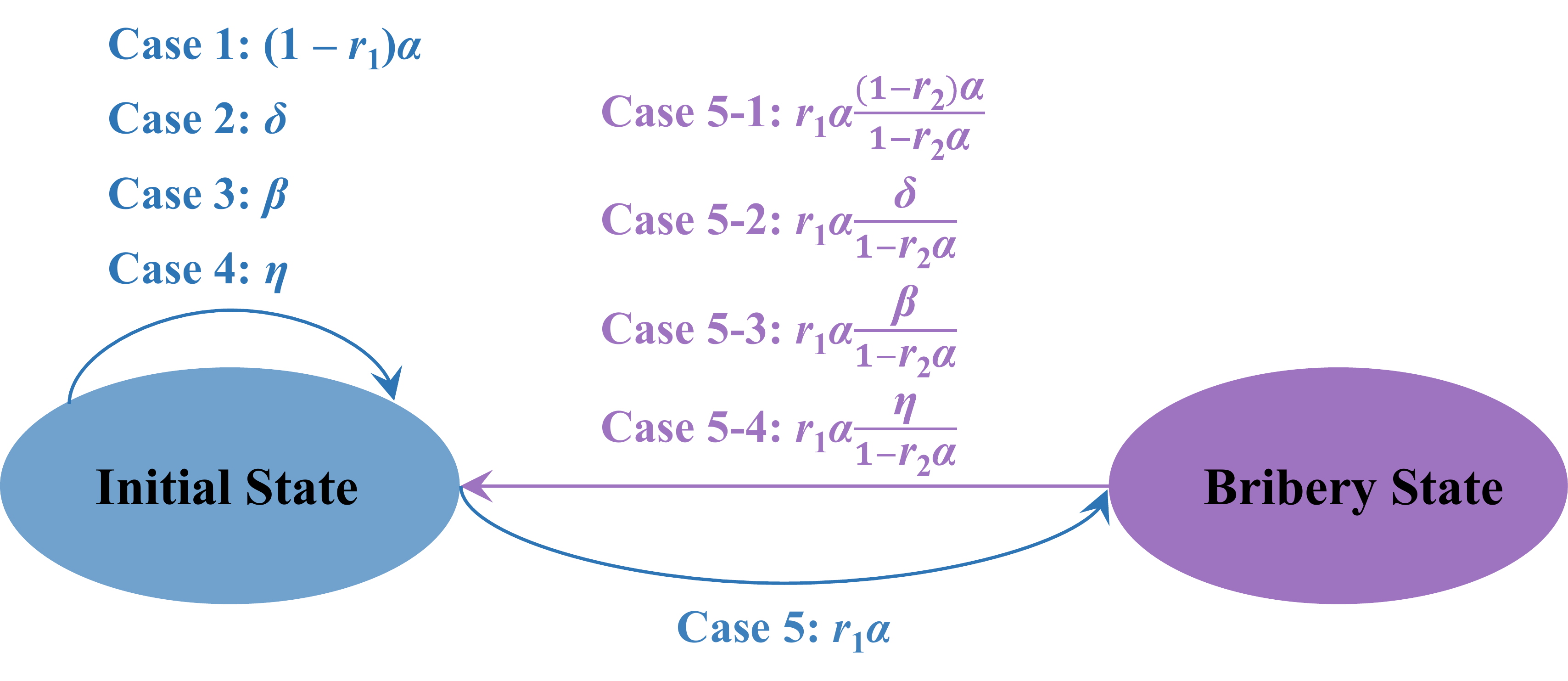}
  \caption{The Markov state transition diagram of BM-PAW attack.}
  \label{Figure}
\end{figure}

In comparison to the PAW attack, our one-pool BM-PAW framework adopts several additional parameters, which are intorduced in Appendix~\ref{apx1}. Specifically, BM-PAW comprises four distinct participant types: the attacker, the victim pool, the target pool, and other miners. We denote the total mining power of the target pool, which consists of miners pre-bribed by the attacker, as $\eta$. Notably, the mining power of other miners, denoted as $\delta$, is calculated as $1 - \alpha - \beta - \eta$, where $\alpha$, $\beta$, and $\eta$ represent the mining power of the attacker, the victim pool, and the target pool, respectively. It is important to clarify that the total mining power of the victim pool ($\beta$) excludes the attacker's infiltration mining power. The Markov state transition diagram of BM-PAW attack is shown in Figure \ref{Figure}.

Prior to power adjustment (Case 5), the attacker allocates $(1-r_1)\alpha$ mining power for honest mining and $r_1 \alpha$ for infiltration mining. Post-adjustment, the attacker employs $(1-r_2)\alpha$ and $r_2 \alpha$ for honest and infiltration mining, respectively. The parameter $c$ is associated with the PoW-based blockchain network topology and the attacker's network capabilities, and its value can be derived using the methodology outlined in the FAW attack.

It is crucial to recognize that, due to the availability of bribe money to any miner selecting the attacker's FPoW, our BM-PAW model may involve multiple targets. In this analysis, $\eta$ represents a single target, as we focus on optimizing the strategy for one target. The involvement of additional targets is contingent upon the bribe money offered, denoted as $\varepsilon_1$ and $\varepsilon_2$, with a greater amount of bribe money attracting more targets.

The relationship between $c$ and $\gamma$ varies based on whether the target accepts or denies the bribe money. Specifically, when the target accepts the bribe (Cases 5-2 and 5-4), the relationship is given by $c_{Case~5 \text{-} 2} =\frac{  \left( 1 - r_{2} \right)\alpha + \eta + \beta + \gamma\delta}{ 1 - r_{2}\alpha }$ and $c_{Case~5 \text{-} 4}=1$, respectively. Conversely, if the target denies the bribe (Cases 5-2' and 5-4'), the relationships are defined as $c_{Case~5 \text{-} 2^{\prime}} =\frac{  \left( 1 - r_{2} \right)\alpha + \beta + \gamma\left(\delta + \eta \right)}{ 1 - r_{2}\alpha }$ and $c_{Case~5 \text{-} 4^{\prime}} =\frac{  \left( 1 - r_{2} \right)\alpha + \beta + \gamma\delta}{ 1 - r_{2}\alpha }$.

The distinction between bribe money $\varepsilon_1$ in Case 5-2 and $\varepsilon_2$ in Case 5-4 arises due to the target's decision to accept or deny the bribe, which is influenced by the specific branch in question. For instance, in Case 5-2, the target may accept even a minimal positive proportion of reward ($\varepsilon_1$). However, in Case 5-4, the target will only accept the bribe if the amount offered exceeds the potential reward loss incurred by accepting it.

In the context of the PAW attack, we contemplate a scenario where the adversary initially partitions their mining power into two components: innocent mining and infiltration mining, within a single victim pool. The adversary possesses the capability to modulate the allocation ratio of their mining power upon discovering a valid FPoW through infiltration mining. A notable distinction between the BM-PAW attack and the traditional PAW attack lies in the fact that the adversary in the BM-PAW attack offers bribe money to the target miners, thereby enhancing the likelihood of their FPoW being selected as the main chain in a forking competition. To elaborate further, there exist five conceivable scenarios when a valid block, specifically an FPoW, is discovered:

\begin{itemize}
    \item [$\bullet$] \noindent\textbf{Case 1.} Found by the attacker’s innocent pool. The attacker will immediately propagate it and obtain the block reward (coin-base reward) in this case.
    \item [$\bullet$] \noindent\textbf{Case 2.} Found by other miners who are none of attacker, target pool and victim pool. The attacker will accept it and continue mining the next block. He cannot get any reward in this case.
    \item [$\bullet$] \noindent\textbf{Case 3.} Found by the victim pool which is none of the attacker’s infiltration mining power. The attacker will accept it and continue mining the next block. He obtains the share reward from this pool in this case.
    \item [$\bullet$] \noindent\textbf{Case 4.} Found by the target pool. This case is similar to Case 2, in which the attacker will accept it and continue mining the next block. He earns nothing in this case.
    \item [$\bullet$] \noindent\textbf{Case 5.} Found by the attacker’s infiltration pool. The attacker will withhold the block and re-adjust his mining power. Meanwhile, there are further four subcases when another valid block/FPoW is found:

\begin{itemize}
    \item \noindent\textbf{Case 5-1.} Found by the attacker’s innocent pool. The attacker will discard the withheld FPoW and propagate the new one. He obtains the block reward in this case.
    \item \noindent\textbf{Case 5-2.} Found by other miners who are none of attacker, target pool and victim pool. The attacker will swiftly submit the previously withheld FPoW to the manager of the victim pool. A blockchain fork occurs when the manager distributes their own FPoW across the network. Simultaneously, the attacker attempts to entice the target miners to mine on the attacker's FPoW by offering an incentive. If the target decides to accept this enticement and mine on the attacker's FPoW, they will receive a bribe paid by the attacker. Alternatively, the target may adopt an honest mining strategy, akin to other miners. It is noteworthy that the target will opt to accept the bribe even if the attacker allocates an extremely small positive proportion of the reward, denoted as $\varepsilon_1$. In this instance, the attacker obtains the forking reward. It is important to acknowledge that the nature of the forking reward differs depending on whether the target accepts or declines the bribe, which will be discussed in greater detail subsequently.
    \item \noindent\textbf{Case 5-3.} Found by the victim pool which is none of the attacker’s infiltration mining power. The attacker will accept it and discard the withheld FPoW. He earns the share reward from this pool in this case.
    \item \noindent\textbf{Case 5-4.} Found by the target pool. The attacker will immediately submit the withheld FPoW to the victim pool manager. Meanwhile, the attacker lures the target to discard his FPoW. The target will get the bribe money paid by the attacker if he chooses to accept the lure and mine on attacker’s FPoW. Otherwise, he will propagate his FPoW and forking occurs. The target may accept bribe money only when the bribe money paid by the attacker is greater than the reward loss suffered by the target in accepting the bribe money. The attacker gets the share reward (the target accepting the bribe money) or forking reward (the target denying the bribe money) in this case.
\end{itemize}
\end{itemize}

\begin{figure}[t]
  \centering
  \includegraphics[width=1.0\linewidth]{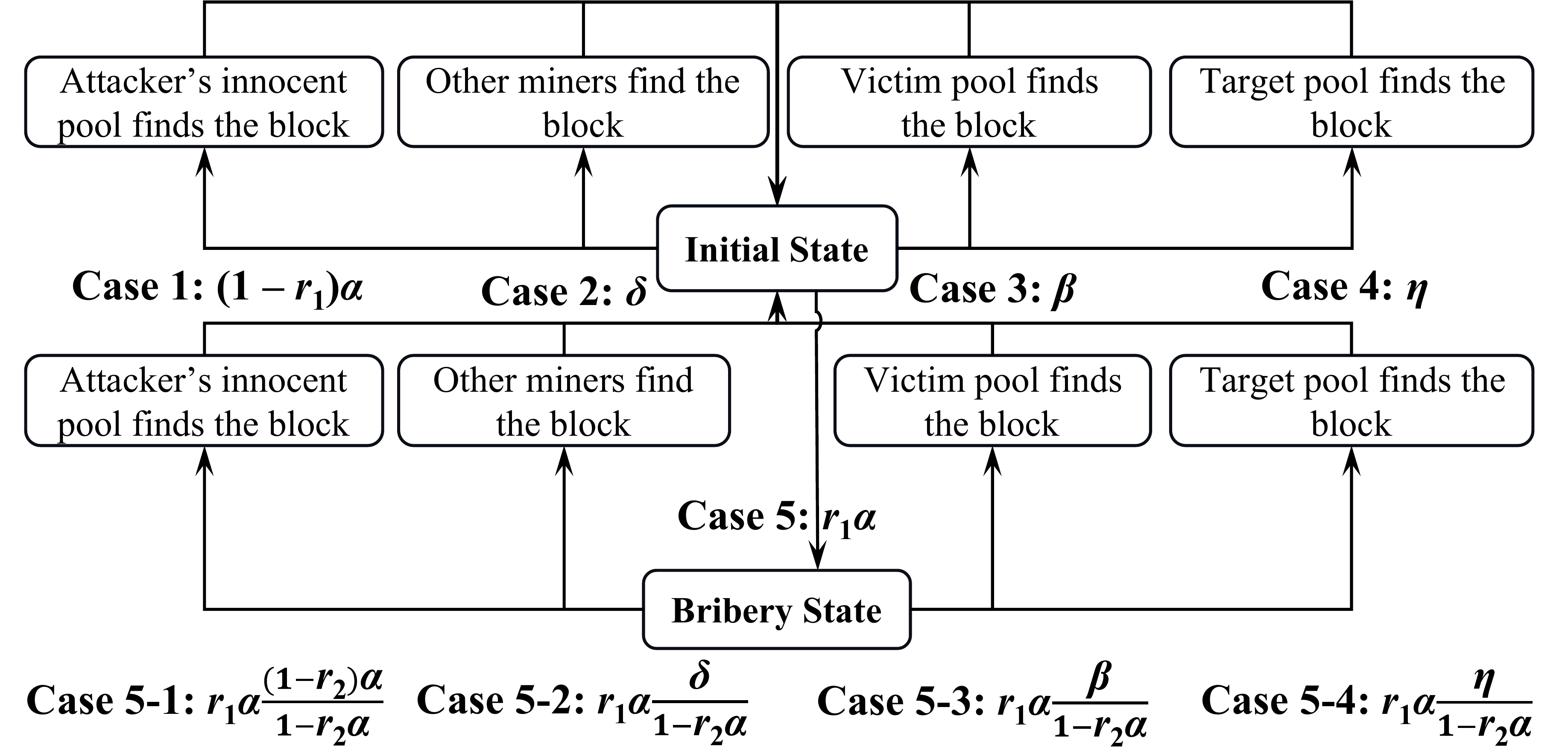}
  \caption{The workflow of the overall BM-PAW attack.}
  \label{Figure n}
\end{figure}

In general, the mining strategy employed in the BM-PAW attack shares similarities with the PAW attack. However, a key distinction lies in the introduction of an additional participant class in the BM-PAW attack: the target miners, which complicates the strategic scenarios. Specifically, we introduce Case 4 during the phase preceding the adjustment of mining power. In this case, if a valid block is discovered by the target pool, the attacker will accept it and proceed to mine the subsequent block. Furthermore, we revise Case 5-2 and introduce Case 5-4 during the phase following the reallocation of mining power. During the forking competition stage, the attacker attempts to persuade the target miners to mine on the attacker's FPoW. In Case 5-2, where the fork occurs between the attacker's FPoW and an FPoW from other miners, the target will accept the bribe even if the attacker offers only an extremely small positive proportion of the reward, denoted as $\varepsilon_1$. Conversely, in Case 5-4, where the fork is between the attacker's FPoW and the target's FPoW, the target may only accept the bribe if the proportion of the reward, denoted as $\varepsilon_2$, is sufficiently large. Additionally, we will demonstrate that the bribe serves as an incentive for the attacker to adjust their attack parameters ($r_1$, $r_2$, $\varepsilon_1$ and $\varepsilon_1$) to maximize their profits. The workflow of the overall BM-PAW attack can be found in Figure \ref{Figure n}.

\subsection{Reward Analysis}
Before analysis, we first define the additional entities in Appendix~\ref{apx2}.

Based on a comprehensive analysis of each aforementioned scenario, we can deduce that the probability of the attacker falling into Case 1 is $(1-r_1)\alpha$; the probability for Case 2 is $\delta$; the probability for Case 3 is $\beta$; the probability for Case 4 is $\eta$; and the probability for Case 5, prior to mining power adjustment, is $r_1 \alpha$. It is noteworthy that the sum of these probabilities, as expected, totals to 1.

Subsequently, after adjusting the mining power, the occurrence probabilities for scenarios 5-1, 5-2, 5-3, and 5-4 are respectively calculated as $r_1 \alpha\cdot\frac{(1-r_2)\alpha}{1-r_2 \alpha}$, $r_1 \alpha\cdot\frac{\delta}{1-r_2 \alpha}$, $r_1 \alpha\cdot\frac{\beta}{1-r_2 \alpha}$, and $r_1 \alpha\cdot\frac{\eta}{1-r_2 \alpha}$. It is important to observe that the cumulative probability of these four scenarios equals $1-r_2 \alpha$, given that infiltration mining ceases to propagate any FPoW post-adjustment of mining power.

When the target accepts the bribe money, we can derive the reward of a BM-PAW attacker $R_a^{BM-PAW^\prime}$ as follows:
\begin{equation}\label{eq1}
R_{a}^{(BM - PAW)}\left( r_{1},r_{2} \right) = R_{a}^{IMR} + R_{a}^{SR} + R_{a}^{FR} - R_{a}^{BM}
\end{equation}

\begin{figure}[t]
  \centering
  \subfloat[]
  {\includegraphics[width=0.45\textwidth]{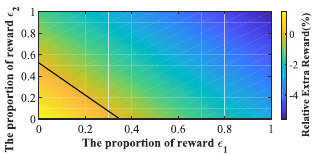}\label{fig:figure1}}
  \quad
  \subfloat[]
  {\includegraphics[width=0.45\textwidth]{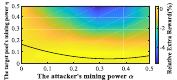}\label{fig:figure2}}
  \quad
  \caption{The attacker’s relative extra reward $RER_a^{BM-PAW,PAW}$ in different scenarios.
  }
\end{figure}

We can get the optimal infiltration mining power $r_1$, $r_2$ and $\overline{r}$ respectively since the proportion of reward as bribe money $\varepsilon_1$ and $\varepsilon_2$ are pre-defined. Therefore, the attacker can adjust his infiltration mining power optimally to maximize his profit.

Clearly, BM-PAW downgrades to PAW when the target denies the bribe money. Therefore, we can further derive the reward of a PAW attacker $R_a^{PAW}$ as follows:
\begin{equation}\label{eq2}
R_{a}^{PAW}\left( r_{1},r_{2} \right) = R_{a}^{IMR} + R_{a}^{SR} + R_{a}^{FR^{\prime}}
\end{equation}

The attacker’s reward is derived by bribe money $R_a^{BM}$, innocent mining reward $R_a^{IMR}$, share reward $R_a^{SR}$ and forking reward $R_a^{FR}$ (the target accepting the bribe money) or $R_a^{FR^\prime}$ (the target denying the bribe money). More specifically, we have two cases for innocent mining reward: Case 1 and Case 5-1. The attacker’s innocent mining reward $R_a^{IMR}$ is $(1-r_1 )\alpha+r_1 \alpha\cdot\frac{(1-r_2) \alpha}{1-r_2 \alpha}$. The share reward is related to two cases: Case 3 and Case 5-3. The attacker’s share reward $R_a^{SR}$ is $\beta\cdot\frac{r_1 \alpha}{r_1 \alpha+\beta}+r_1 \alpha\cdot\frac{\beta}{1-r_2 \alpha}\cdot\frac{\overline{r}\alpha}{\overline{r}\alpha+\beta}$. For forking reward, we have two cases: Case 5-2 and Case 5-4. The attacker’s forking reward $R_a^{FR}$ is $c_{Case~5 \text{-}2}r_1\alpha \cdot \frac{\delta}{ 1 - r_2\alpha} \cdot\frac{ \overline{r}\alpha}{ \overline{r}\alpha + \beta }$ in Case 5-2 and $c_{Case~5 \text{-}4}r_1\alpha \cdot \frac{\eta}{ 1 - r_2\alpha} \cdot\frac{ \overline{r}\alpha}{ \overline{r}\alpha + \beta }$ in Case 5-4 when the target chooses to accept the bribe money. Furthermore, the attacker’s bribe money $R_a^{BM}$ consists of two parts, namely $\varepsilon_1 c_{Case~5 \text{-}2} r_1 \alpha\cdot\frac{\delta}{ 1 - r_2\alpha} \cdot\frac{ \overline{r}\alpha}{ \overline{r}\alpha + \beta }$ in Case 5-2 and $\varepsilon_2 c_{Case~5 \text{-}4}r_1\alpha \cdot \frac{\eta}{ 1 - r_2\alpha} \cdot\frac{ \overline{r}\alpha}{ \overline{r}\alpha + \beta }$ in Case 5-4. Once the target denies the bribe money, the attacker’s forking reward $R_a^{FR^\prime}$ is $c_{Case~5 \text{-}2^{\prime}}r_1\alpha \cdot \frac{\delta}{ 1 - r_2\alpha} \cdot\frac{ \overline{r}\alpha}{ \overline{r}\alpha + \beta }$ in Case 5-2 and $c_{Case~5 \text{-}4^{\prime}}r_1\alpha \cdot \frac{\eta}{ 1 - r_2\alpha} \cdot\frac{ \overline{r}\alpha}{ \overline{r}\alpha + \beta }$ in Case 5-4. In summary, $R_a^{BM-PAW}$ is derived by separating the attacker’s reward into $R_a^{IMR}$, $R_a^{SR}$, $R_a^{FR}$, and $R_a^{BM}$ while $R_a^{PAW}$ is related to $R_a^{IMR}$, $R_a^{SR}$ and $R_a^{FR^\prime}$. Therefore, we can derive the reward of a BM-PAW attacker $R_a^{BM-PAW}$ in Equation (\ref{eq1}) and the reward of a PAW attacker $R_a^{PAW}$ in Equation (\ref{eq2}).

Then, we can derive the attacker’s extra reward $R_a$ as follows:

\begin{equation}\label{eq3}
\begin{aligned}
R_{a}( r_{1},r_{2}) &= R_{a}^{BM - PAW}( r_{1},r_{2} ) - R_{a}^{PAW}( r_{1},r_{2})= R_{a}^{FR} - R_{a}^{BM} - R_{a}^{FR^{\prime}} \\
&= r_{1}\alpha\left(
\begin{array}{lcl}
\left(( 1 - \varepsilon_{1})c_{Case~5 \text{-}2} - c_{Case~5 \text{-}2^{\prime}}\right)\frac{\delta}{1 - r_{2}\alpha}\\
+\left(( 1 - \varepsilon_{2} )c_{Case~5 \text{-}4} - c_{Case~5 \text{-}4^{\prime}}\right)\frac{\eta}{ 1 - r_{2}\alpha}
\end{array}
\right)
\frac{\overline{r}\alpha}{\overline{r}\alpha + \beta}
\end{aligned}
\end{equation}

\textit{How to adjust the attacker’s infiltration mining power that could maximize his rewards?} Based on the attacker’s system reward in BM-PAW, we formalize the optimal strategy of mining power adjusting by finding the best proportion of attacker’s mining power ($\hat{r}_1$ and $\hat{r}_2$) to maximize his expected reward ($R_a^{BM-PAW^\prime} (r_1,r_2)$):

\begin{equation}\label{eq4}
\begin{split}
&{\arg{\max\limits_{r_{1},r_{2}}{R_{a}^{{BM - PAW}^\prime}\left( {r_{1},~r_{2}} \right)}}},\\
&s.t.~~0 \leq r_{1} \leq 1,~0 \leq r_{2} \leq 1.    
\end{split}
\end{equation}

We can use Lagrange multipliers to solve Equation (\ref{eq3}). Specifically, we rewrite Equation (\ref{eq4}) as follows:

\begin{equation}
\begin{aligned}
\arg&\min\limits_{r_{1},r_{2}}{- R_{a}^{{BM - PAW}^\prime}\left( {r_{1},~r_{2}} \right)},\\
s.t.~~&g_{1}\left( {r_{1},r_{2}} \right) = - r_{1} \leq 0;\\
&g_{2}\left( {r_{1},r_{2}} \right) = r_{1} - 1 \leq 0;\\
&g_{3}\left( {r_{1},r_{2}} \right) = - r_{2} \leq 0;\\
&g_{4}\left( {r_{1},r_{2}} \right) = r_{2} - 1 \leq 0;
\end{aligned}
\end{equation}

As expected, the objective function $-R_a^{BM-PAW^\prime}(r_1,r_2)$ is a convex function when $r_1, r_1\in\left[0,1\right]$ as the Hessian matrix is positive definite. Hence, we can find the optimal $\hat{r}_1$ and $\hat{r}_2$ by solving the KKT conditions, which are shown in Table \ref{table1}.

\begin{table}[t]
\setlength{\tabcolsep}{10pt}
\centering
\caption{The attacker’s optimal infiltration mining power $\hat{r}_1$ and $\hat{r}_2$ in BM-PAW in different scenarios. The values $\hat{r}_1(\hat{r}_2)$, $\hat{r}_1$, $\hat{r}_2$ indicate the optimal infiltration mining power before and after the mining power adjusting respectively.}\label{table1}
\begin{tabular}{@{}ccccc@{}}
\toprule
$\beta$ &  $\alpha$=0.1 & $\alpha$=0.2 &$\alpha$=0.3 &$\alpha$=0.4  \\ \midrule
0.1	& 0.1404(0.9985) &  0.1222(0.6787) & 0.1254(0.4903) & 0.1372(0.3938)\\
0.2	& 0.3039(0.9996) & 0.2844(0.9993) & 0.2928(0.7998) & 0.3271(0.6334)\\
0.3	& 0.2745(0.6771) & 0.3079(0.3098) & 0.3691(0.1313) & 0.4791(0.0002)\\
\bottomrule
\end{tabular}
\end{table}


Meanwhile, we can derive the following theorem with the optimal $\hat{r}_1$ and $\hat{r}_2$.  Detailed proofs of Theorems \ref{theorem1} and \ref{theorem2} are provided in \cite{nihao}.

\begin{theorem}\label{theorem1}
A BM-PAW attacker can always earn more reward than honest mining, and the reward of a BM-PAW attacker has a lower bound defined by the reward from an FAW attack.
\end{theorem}

\begin{theorem}\label{theorem2}
When $\varepsilon_1$ and $\varepsilon_2$ satisfy $\delta c_{Case~5 \text{-} 2} \cdot \varepsilon_{1} + \eta c_{Case~5 \text{-} 4} \cdot \varepsilon_{2} < \delta( c_{Case~5 \text{-} 2} -$

\noindent$c_{{Case~5 \text{-} 2}^\prime} ) + \eta\left( {c_{Case~5 \text{-} 4} - c_{{Case~5 \text{-} 4}^\prime}} \right)$, a BM-PAW attacker can always obtain more reward than a PAW attacker.
\end{theorem}

\begin{figure}[t]
  \centering
  \subfloat[]
  {\includegraphics[width=0.45\textwidth]{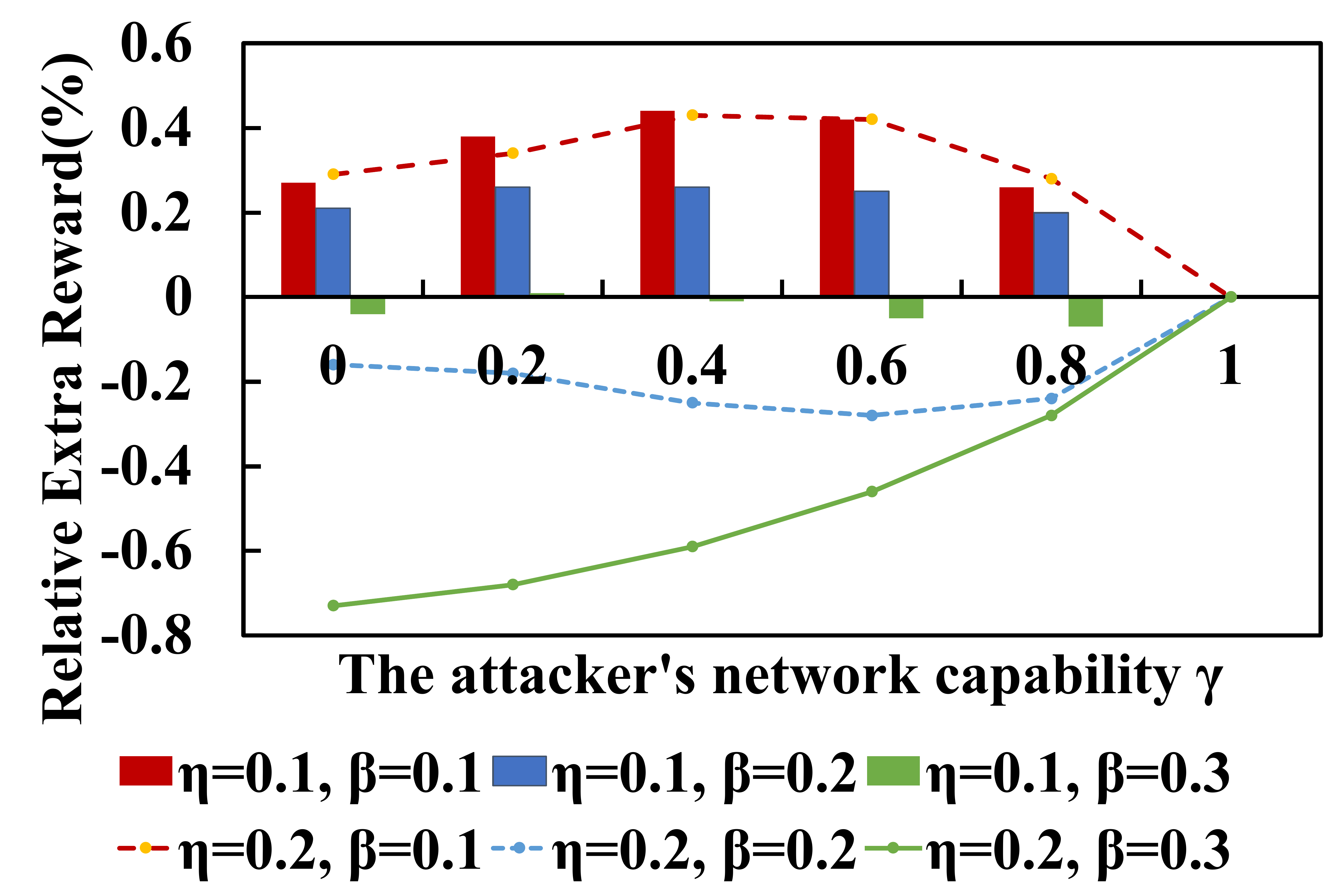}\label{fig:figure3}}
  \quad
  \subfloat[]
  {\includegraphics[width=0.45\textwidth]{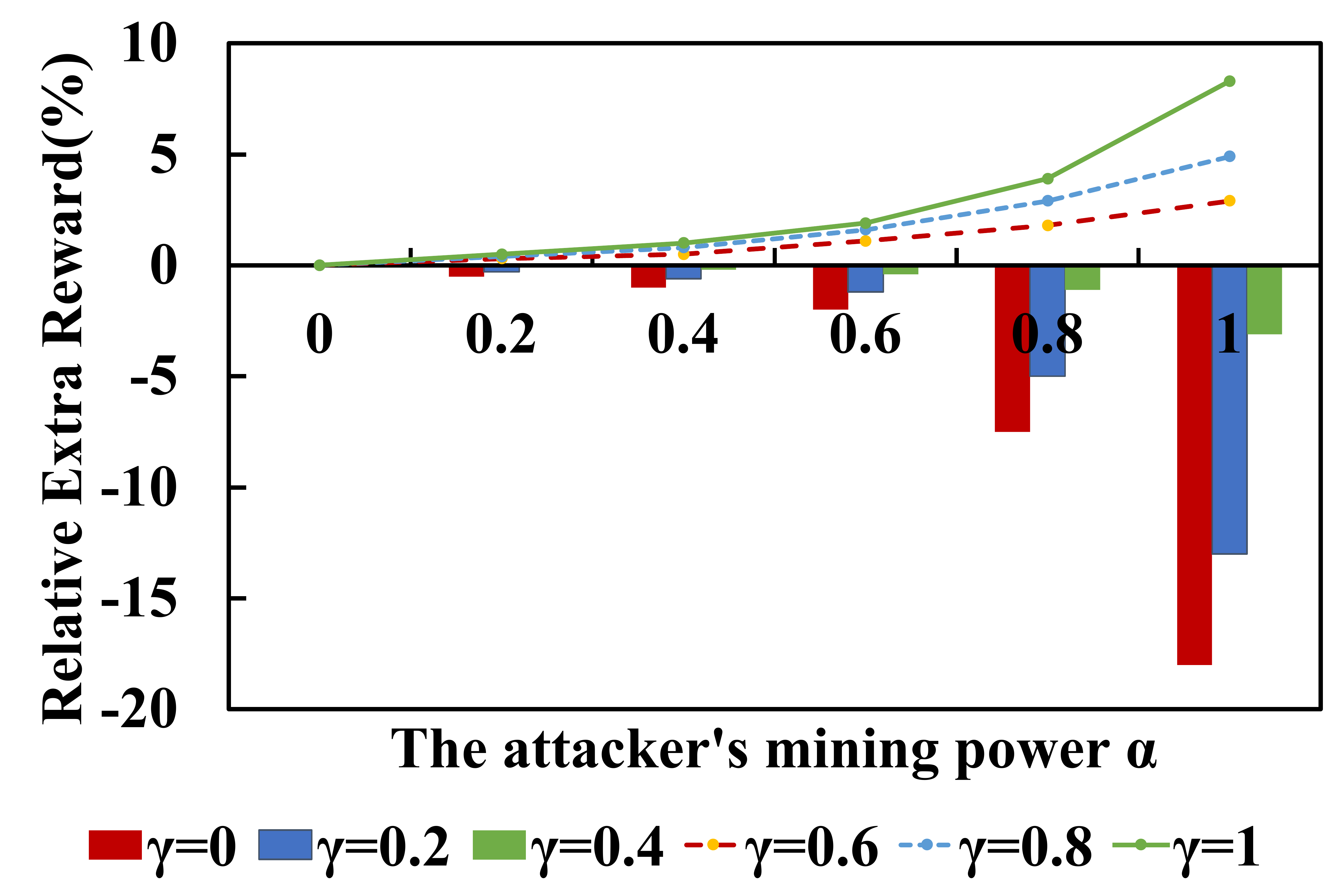}\label{fig:figure4}}
  \quad
  \caption{The attacker’s relative extra reward $RER_a^{BM-PAW,PAW}$ in different parameters.
  }
\end{figure}

\subsection{Implementation and Evaluation}
We adopt expected relative extra reward (RER) to evaluate BM-PAW attack, which can be expressed as ${RER}_{\tau}^{S_{1},S_{2}} = \frac{R_{\tau}^{S_{1}} - R_{\tau}^{S_{2}}}{R_{\tau}^{S_{2}}}$, where $\tau$ represents the participant role (i.e., the attacker ($a$), the target ($t$) and etc.), and $S$ indicates different strategies (i.e., BM-PAW, PAW, etc.). We implement a Monte Carlo simulator in MATLAB to verify the accuracy of our theoretical analysis of BM-PAW attack. We run it over $10^9$ rounds to derive a BM-PAW attacker’s reward when $\alpha=0.2$, $\beta=\{0.1,0.2,0.3\}$ and $\eta=\{0.1,0.2\}$. Then we calculate the attacker’s RER during the process.

First, consider a specific case: the attacker has mining power $\alpha=0.2$, the victim pool has mining power $\beta=0.2$, the target pool has mining power $\eta=0.2$, and the ratio of other miners that select the attacker’s FPoW $\gamma=0.5$. $RER_a^{BM-PAW,PAW}$ with varying $\varepsilon_1$ and $\varepsilon_2$ is depicted in Fig. \ref{fig:figure1}. The left side of the solid line indicates that the BM-PAW attacker can obtain more reward than in PAW. The BM-PAW attacker’s RER increases when the proportion of reward $\varepsilon_1$ and $\varepsilon_2$ decrease, since the bribe money is determined by $\varepsilon_1$ and $\varepsilon_2$. More specifically, the smaller $\varepsilon_1$ and $\varepsilon_2$ are, the lower the bribe money is, which is consistent with our reward analysis.

Moreover, considering $\beta=0.2$, the minimum $\varepsilon_1$ and $\varepsilon_2$ when the target pool is willing to accept the bribe money, $RER_a^{BM-PAW,PAW}$ with varying $\alpha$ and $\eta$ is shown in Fig. \ref{fig:figure2}. The lower part of the solid line indicates that BM-PAW is the dominant strategy compared with PAW, which can bring higher profit for the BM-PAW attacker. In addition, the attacker’s RER increases when the victim pool’s mining power decreases, since the attacker can get more share reward from target pool. Meanwhile, the attacker has the motivation to launch BM-PAW attack when the target pool possesses little mining power regardless of his own mining power, which is more profitable.

Furthermore, we take into account $\alpha=0.2$, $\eta=\{0.1,0.2\}$, $\beta=\{0.1,0.2,0.3\}$, the minimum $\varepsilon_1$ and $\varepsilon_2$, $RER_a^{BM-PAW,PAW}$ with varying $\gamma$ is shown in Fig. \ref{fig:figure3}. The attacker’s RER is related to his network capability. He can earn higher profit when the victim pool infiltrated by the attacker has less mining power, regardless of his network capability, since the attacker can earn more share reward from target pool. Therefore, the pool with lower mining power is more vulnerable to attack.

Finally, Fig. \ref{fig:figure4} represents $RER_a^{BM-PAW,PAW}$ with varying $\alpha$, and constant $\beta=0.2$, $\eta=0.2$, the minimum $\varepsilon_1$ and $\varepsilon_2$, $\gamma=\{0:0.1:1\}$, which indicates that the higher attacker’s network capability cannot always obtain higher RER. The reason is that the attacker’s network capability will affect both BM-PAW and PAW simultaneously.

\section{Bribe Money Pricing}
\subsection{Overview}
We assume that there exists a secret protocol with the bribe money ratio $\varepsilon_1$ in Case 5-2 as well as $\varepsilon_2$ in Case 5-4 to launch the BM-PAW attack between the BM-PAW attacker and the target pool. Here, we focus on analyzing the constraint condition of $\varepsilon_1$ and $\varepsilon_2$ in order to bring about the BM-PAW attack effective and profitable. We let $R_t^{BM-PAW}$ be the target pool’s total reward in BM-PAW (i.e., the target accepting the bribe money) and $R_t^{PAW}$ be the target pool’s total reward in PAW (i.e., the target denying the bribe money) respectively. When the attacker cannot afford enough bribe money to motivate the target pool to accept it, the BM-PAW attack downgrades to the PAW attack. Meanwhile, the attacker’s reward should be improved in the BM-PAW attack as well after paying bribe money to the target. Otherwise, it is more profitable for attacker to exploit the PAW attack. We can get the reward of the target pool in the PAW attack denoted by $R_t^{PAW}$ which will be adopted as the baseline of the bribe money ratio $\varepsilon_1$ and $\varepsilon_2$ with the same parameters. What is the most significant is that the minimum bribe money must be sufficient to guarantee the reward of the target in BM-PAW is greater than in PAW and the maximum bribe money must guarantee the reward of the attacker in BM-PAW is larger than in PAW, namely $R_{t}^{BM - PAW} > R_{t}^{PAW}$ and $R_{a}^{BM - PAW} > R_{a}^{PAW}$.

\subsection{Reward Analysis}
First, when the target pool accepts the bribe money, we can obtain the target pool’s reward $R_t^{BM-PAW}$ in the BM-PAW attack as follows:
\begin{equation}
\begin{aligned}
&R_{t}^{BM - PAW}\left( {r_{1},~r_{2}} \right) = \eta + r_{1}\alpha\left( {\frac{\delta}{1 - r_{2}\alpha} + \frac{\eta}{1 - r_{2}\alpha}} \right)\frac{\eta}{1 - r_{2}\alpha} \\
&+ r_{1}\alpha\left( {{\varepsilon_{1}c}_{Case~5 \text{-} 2} \cdot \frac{\delta}{1 - r_{2}\alpha} + \varepsilon_{2}c_{Case~5 \text{-} 4} \cdot \frac{\eta}{1 - r_{2}\alpha}} \right)\frac{\overset{-}{r}\alpha}{\overset{-}{r}\alpha + \beta}
\end{aligned}
\end{equation}

\begin{figure}[t]
  \centering
  \subfloat[]
  {\includegraphics[width=0.45\textwidth]{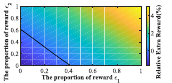}\label{fig:figure5}}
  \quad
  \subfloat[]
  {\includegraphics[width=0.45\textwidth]{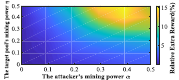}\label{fig:figure6}}
  \quad
  \caption{The target pool’s relative extra reward $RER_t^{BM-PAW,PAW}$ in different scenarios.
  }
\end{figure}

Furthermore, when the target pool denies the bribe money, we derive the target pool’s reward $R_t^{PAW}$ in the PAW attack as follows:
\begin{equation}
\begin{aligned}
R_{t}^{PAW}\left( {r_{1},~r_{2}} \right) &= \eta + r_{1}\alpha \cdot \frac{\delta}{1 - r_{2}\alpha} \cdot \frac{\eta}{1 - r_{2}\alpha} \\
&+ r_{1}\alpha \cdot \frac{\eta}{1 - r_{2}\alpha}\left( {1 - c_{Case~{5 \text{-} 4}^\prime} + \frac{\eta}{1 - r_{2}\alpha}} \right)
\end{aligned}
\end{equation}

Therefore, we can get the target pool’s extra reward $R_t$ as follows:
\begin{equation}
\begin{aligned}
R_{t}\left( {r_{1},~r_{2}} \right) &= R_{t}^{BM - PAW}\left( {r_{1},~r_{2}} \right) - R_{t}^{PAW}\left( {r_{1},~r_{2}} \right) \\
&= r_{1}\alpha\left( {{\varepsilon_{1}c}_{Case~5 \text{-} 2} \cdot \frac{\delta}{1 - r_{2}\alpha} + \varepsilon_{2}c_{Case~5 \text{-} 4} \cdot \frac{\eta}{1 - r_{2}\alpha}} \right)\frac{\overset{-}{r}\alpha}{\overset{-}{r}\alpha + \beta} \\
&- r_{1}\alpha \cdot \frac{\eta}{1 - r_{2}\alpha}\left( {1 - c_{Case~{5 \text{-} 4}^\prime}} \right)
\end{aligned}
\end{equation}

\textit{How to choose the proper bribe money to achieve incentive compatibility between the attacker and the target pool?} Below, we present Theorem \ref{thm3} to address the aforementioned questions. Detailed proofs of Theorems \ref{thm3} and \ref{thm4} are provided in \cite{nihao}.
\begin{theorem}\label{thm3}
When $\varepsilon_1$ and $\varepsilon_2$ satisfy $\delta c_{Case~5 \text{-} 2} \cdot \varepsilon_{1} + \eta c_{Case~5 \text{-} 4} \cdot \varepsilon_{2} > \eta( 1 -$

\noindent$c_{Case~{5 \text{-} 4}^\prime} )\frac{\overset{-}{r}\alpha + \beta}{\overset{-}{r}\alpha}$, the target pool can always get higher reward in the BM-PAW attack than in the PAW attack.
\end{theorem}

\begin{theorem}\label{thm4}
Once launching BM-PAW, both the BM-PAW attacker and the target pool can get higher reward than in the PAW attack when the attacker pays proper bribe money and the target accepts it.    
\end{theorem}

The result of Theorem \ref{thm4} guarantees that a rational attacker is willing to launch BM-PAW attack and a greedy target pool has motivation to accept the bribe money and join it. Furthermore, once the attacker decides to execute the BM-PAW attack with pre-determined $\varepsilon_1$ and $\varepsilon_2$, it is motivated for him to adjust the optimal attack parameters (i.e., $r_1$ and $r_2$) to maximize his extra reward. Notice that a PAW attacker adopts a similar approach to determine the optimal $r_1$ and $r_2$.

\subsection{Implementation and Evaluation}
We further use a Monte Carlo simulator verify our analysis. We run it over $10^9$ rounds when $\alpha=0.2$, $\eta=\{0.1,0.2,0.3\}$ and $\beta=\{0.1,0.2,0.3\}$. Then we calculate the target pool’s RER during the process. 

First, Fig. \ref{fig:figure5} shows $RER_t^{BM-PAW,PAW}$ with varying $\varepsilon_1$ and $\varepsilon_2$, and constant $\alpha=0.2$, $\beta=0.2$, $\eta=0.2$ and $\gamma=0.5$. The right side of the solid line indicates that the target pool accepting the bribe money can obtain more reward than denying. The target’s RER increases with $\varepsilon_1$ and $\varepsilon_2$ increasing, since the target will get higher reward with the attacker paying more bribe money. Therefore, accepting the bribe money is always the optimal strategy for the target when the attacker offers proper bribe money.

\begin{figure}[t]
  \centering
  \subfloat[]
  {\includegraphics[width=0.45\textwidth]{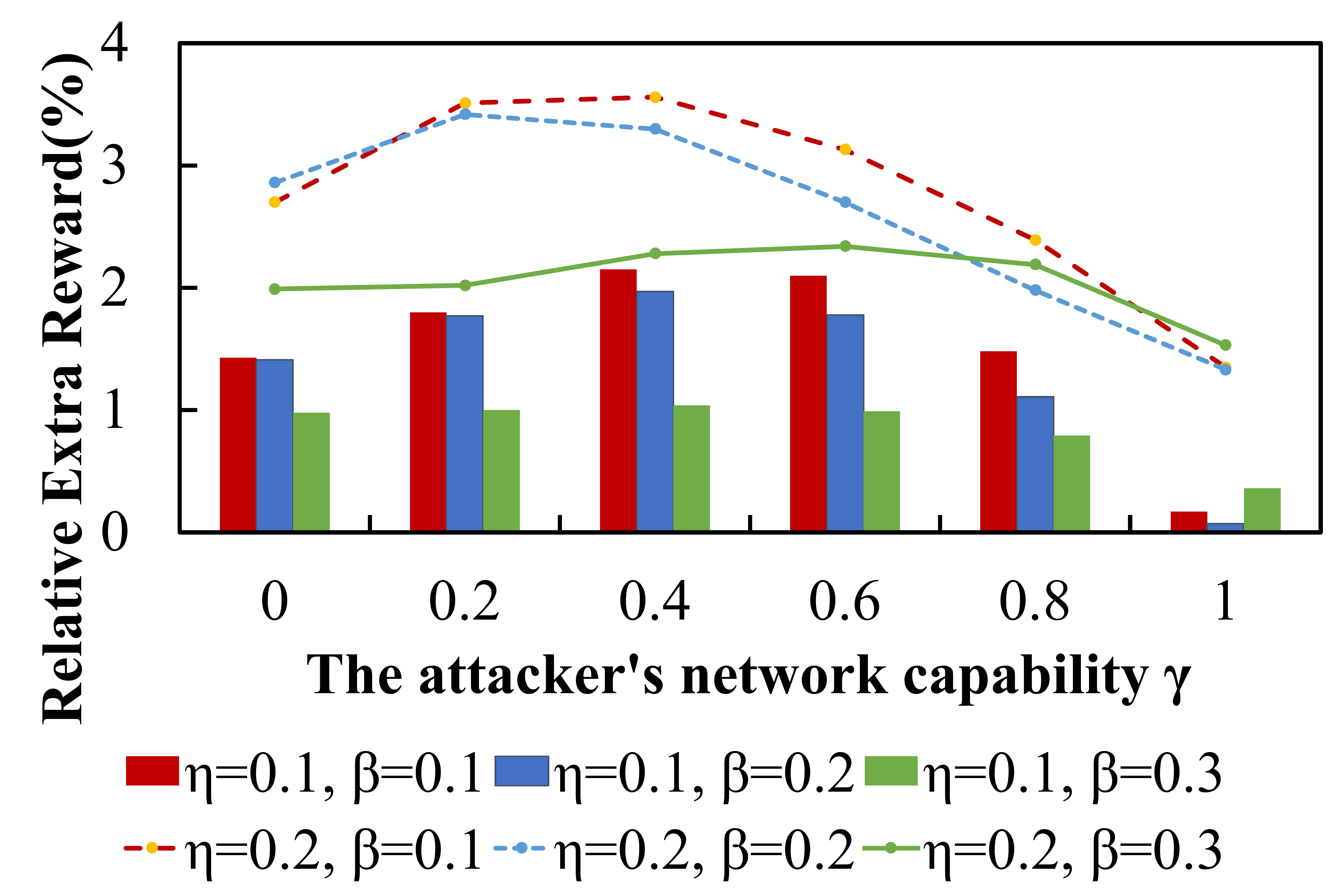}\label{fig:figure7}}
  \quad
  \subfloat[]
  {\includegraphics[width=0.45\textwidth]{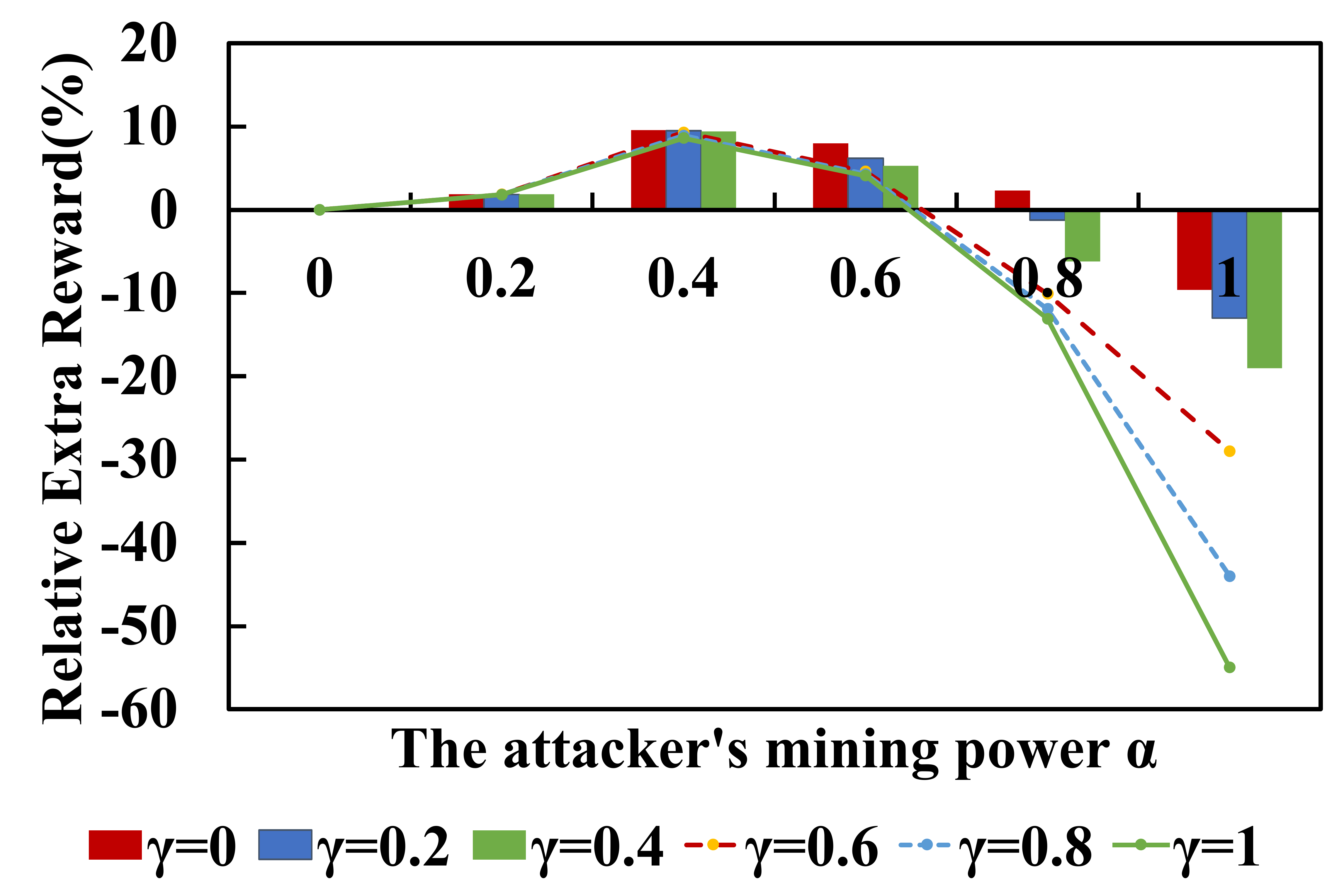}\label{fig:figure8}}
  \quad
  \caption{The target pool’s relative extra reward $RER_t^{BM-PAW,PAW}$ in different parameters.
  }
\end{figure}

Moreover, Fig. \ref{fig:figure6} represents $RER_t^{BM-PAW,PAW}$ with varying $\alpha$ and $\eta$, and constant $\beta=0.2$, $\eta=0.2$, $\gamma=0.5$, and the maximum $\varepsilon_1$ and $\varepsilon_2$. The target pool can always get more reward when the attacker pays the maximum bribe money. Hence, accepting the bribe money is the optimal strategy for the target, regardless of the mining power he possesses.

Furthermore, considering $\alpha=0.2$, $\eta=\{0.1,0.2\}$, $\beta=\{0.1,0.2,0.3\}$, the maximum $\varepsilon_1$ and $\varepsilon_2$, Fig. \ref{fig:figure7} shows $RER_t^{BM-PAW,PAW}$ with varying $\gamma$, which indicates the target has the incentive to accept the bribe money for most cases. Similarly, the target’s RER increases when the victim pool’s mining power decreases, since the target can get more bribe money from the attacker’s share reward.

Finally, Fig. \ref{fig:figure8} depicts $RER_t^{BM-PAW,PAW}$ with varying $\alpha$, and constant $\beta=0.2$, $\eta=0.2$, the maximum $\varepsilon_1$ and $\varepsilon_2$, and $\gamma=\{0:0.1:1\}$. We derive that the higher attacker’s network capability cannot always obtain higher RER for the target since both BM-PAW and PAW are determined by the attacker’s network capability simultaneously.

\section{Two-pool: Miner’s Dilemma}
Based on the analyze of the BM-PAW attack, we find that two mining pools can also exploit the BM-PAW attack against each other. We demonstrate that two-pool BM-PAW game can avoid the “miner’s dilemma”, in which the single winner is determined by the pool size (i.e., the larger pool can win). We define the winning condition as obtaining an extra reward to analyze the two-pool BM-PAW attack game.

We consider a game involving Pool$_1$ and Pool$_2$ with mining power $\alpha^{[i]}$($i\in\{1,2\}$). The Pool$_i$ will distribute the infiltration mining power either $f_{1}^{\lbrack i\rbrack} = r_{1}^{\lbrack i\rbrack}\alpha^{\lbrack i\rbrack}$ (before the mining power adjusting) or $f_{2}^{\lbrack i\rbrack} = r_{2}^{\lbrack i\rbrack}\alpha^{\lbrack i\rbrack}$ (after the mining power adjusting) in his opponent Pool$_{\neg i}$, where $\neg i = 3 - i$ and $i\in\{1,2\}$. Therefore, we further define the additional parameters in Appendix~\ref{apx3}.

\begin{table}[t]
\setlength{\tabcolsep}{8pt}
\caption{Results of a BM-PAW game with varying $\alpha^{[2]}$ and $c$ when $\alpha^{[1]}=0.2$. The left side shows the RER of Pool$_1$, and the right side shows the RER of Pool$_2$.}\label{table2}
\centering
\renewcommand\arraystretch{1}
\newcolumntype{"}{@{\hskip\tabcolsep\vrule width 1pt\hskip\tabcolsep}}
\begin{tabular}{cccc"cccc}
 \bottomrule[1pt]
$\alpha^{[1]}$ & $c=0.2$ & $c=0.6$ & $c=1$ & $\alpha^{[2]}$ & $c=0.2$ & $c=0.6$ & $c=1$\\
\hline
0.1	& 0.9810	& 0.9982	&1.0155	&0.1	&-0.4952	&0.4995	&-0.5038\\
0.2	&-0.0177	&-0.0089	&0	 &0.2	&0.01808	&0.0090	&0\\
0.3	&-0.3491	&-0.3441	&-0.3392	&0.3	&0.5364	&0.5248	&0.5132\\
0.4	&-0.5163	&-0.5134	&-0.5104	&0.4	&1.0678	&1.0550	&1.0424\\
\toprule[1pt]
\end{tabular}
\end{table}







To calculate $R_{a}^{\lbrack i\rbrack}$, we introduce the possible cases (which are parameterized by Pool$_i$) when a valid block (FPoW) is found in a two-pool BM-PAW game, which can be found in Appendix \ref{apx4}.









In a two-pool BM-PAW game, the Pool$_i$’s reward $R_{a}^{\lbrack i\rbrack}$ is derived from bribe money, innocent mining reward, share reward and forking reward.



\textit{What equilibrium will the BM-PAW game of different types of participants eventually reach?} To answer the above question, we quantitatively analyze the reward in a two-pool BM-PAW game under the Nash equilibrium point. To simplify our analysis, we assume $c_{1}^{\lbrack 1\rbrack} = c_{1}^{\lbrack 2\rbrack} = c$ and $c_{2}^{\lbrack 1\rbrack} = c_{2}^{\lbrack 2\rbrack} = c/2$, where $0\leq c \leq1$. Table \ref{table2} shows the results of a two-pool BM-PAW game with varying $\alpha^{[2]}$ and $c$ when $\alpha^{[1]}$=0.2. We find that Pool$_2$ may get more reward when $\alpha^{[2]}$>0.2, which is determined by his network capability. However, when $\alpha^{[1]}<0.2$, Pool$_2$ will always suffer a loss. For Pool$_1$, it still follows the above pattern. Therefore, we can derive that the winning condition is determined by the attacker’s mining power in a two-pool BM-PAW game, which breaks the “miner’s dilemma”.

\section{Dicussion}
We propose three countermeasures to address the emerging threat posed by bribery attacks utilizing monetary incentives. Initially, upon the occurrence of a blockchain fork, honest miners ought to prioritize and select FPoW they detect. As an illustrative example, if a miner detects transaction $T^A$ prior to a fork event that subsequently introduces a branch containing both transactions $T^A$ and $T^B$, the miner should expand the blockchain along the branch incorporating $T^A$. If all miners adhere to this mining strategy, the effectiveness of bribery attacks can be significantly mitigated. However, it is important to acknowledge the impracticality of this assumption due to miners' profit-oriented nature, which may prompt them to opt for an alternative branch containing $T^B$ to maximize their rewards. Despite this limitation, it is noteworthy that as the number of miners adopting this strategy increases, the value of $\gamma$ decreases, resulting in reduced rewards for the attacker and consequently lowering the success probability of their FPoW in achieving consensus.

Secondly, upon detecting a bribery attack, the victim may consider employing a counter-bribery strategy by offering additional monetary incentives to secure higher profits. Typically, any miner who secures a reward on the legitimate main chain, rather than on the attacker's forked branch, can adopt such a counter-bribery approach. It is crucial to note that the victim's expenditure on counter-bribery measures should not exceed the total value of the transaction $T^A$. In the event that the attacker prevails in the competition, the victim stands to lose the entire value of $T^A$. Consequently, the attacker must increase the amount of bribe money offered to the target, rendering the bribery attack economically unviable.

Thirdly, there exists the potential for dynamic shifts in the roles of individual miners or mining pools. Specifically, the identities of the attacker, the victim, and the target pool are subject to continuous change over time. A miner who currently benefits from short-term gains as an attacker may subsequently incur losses in the role of a victim. Consequently, if the immediate rewards derived from bribery money compromise the long-term profit potential, miners will lack the motivation to initiate BM-PAW or to accept bribe money.
\section{Conclusion}
We introduce a novel mining attack, termed BM-PAW, which augments the existing PAW attack by incorporating a bribery strategy. We reveal that the BM-PAW attack can offer higher rewards to both the attacker and the target mining pool compared to the traditional PAW attack. While our primary focus is on the PoW-based blockchain system, the BM-PAW attack has the potential to be adapted for use in other PoW-based cryptocurrency systems. Through equilibrium analysis in a two-pool BM-PAW game scenario, where the winning condition is contingent upon the attacker's mining power, we find that the BM-PAW attacker can evade the classic “miner's dilemma”. We finally propose practical countermeasures to mitigate these novel bribery-based pool attacks, which may offer fresh insights and inspiration to researchers in the field.

\section*{Acknowledgments} 
The authors thank Huan Yan and Xunzhi Chen for the discussion in the early stage of this work. This research has received support from National Key R\&D Program of China (2023YFB2704700), National Natural Science Foundation of China (62472276), Shanghai Committee of Science and Technology, China (24BC3200400), Science and Technology Project of the State Grid Corporation of China (5700-202321603A-3-2-ZN).

%
%
%
\bibliographystyle{splncs04}
%

\appendix

\section{Symbol Definition}
\subsection{}\label{apx1}
We adopt the following parameters to analyze the BM-PAW attack:\\

\begin{itemize}
    \item[$\bullet$] $\alpha$: The total mining power of the attacker;
    \item[$\bullet$] $\beta$: The total mining power of the victim pool;
    \item[$\bullet$] $\eta$: The total mining power of the target pool;
    \item[$\bullet$] $\delta$: The mining power of other miners who are none of attacker, target pool and victim pool;
    \item[$\bullet$] $r_1$: The original infiltration mining power of the attacker as a proportion of $\alpha$ before power adjusting (Case 5) in each round;
    \item[$\bullet$] $r_2$: The reallocated infiltration mining power of the attacker as a proportion of $\alpha$ after power adjusting (Case 5) in each round;
    \item[$\bullet$] $\overline{r}$: The average infiltration mining power of the attacker as a proportion of $\alpha$ between a mining process;
    \item[$\bullet$] $c$: The probability of the attacker’s FPoW will be selected as the main chain in Case 5-2 or Case 5-4.
    \item[$\bullet$] $\gamma$: The ratio of other miners that select the attacker’s FPoW;
    \item[$\bullet$] $\varepsilon_1$: The fraction of reward as bribe money of the attacker willing to pay per block when the target pool chooses to accept bribe money in Case 5-2;
    \item[$\bullet$] $\varepsilon_2$: The proportion of reward as bribe money of the attacker willing to pay per block when the target pool chooses to accept bribe money in Case 5-4.
\end{itemize}

\subsection{}\label{apx2}
We define the following entities to analyze the reward in BM-PAW attack:\\

\begin{itemize}
    \item[$\bullet$] $R_a^{IMR}$: The attacker’s innocent mining reward;
    \item[$\bullet$] $R_a^{SR}$: The attacker’s share reward;
    \item[$\bullet$] $R_a^{FR}$: The attacker’s forking reward when the target accepts the bribe money;
    \item[$\bullet$] $R_a^{FR^\prime}$:	The attacker’s forking reward when the target denies the bribe money;
    \item[$\bullet$] $R_a^{BM}$: The attacker’s bribe money;
    \item[$\bullet$] $R_a^{BM-PAW^\prime}$:	The attacker’s system reward in BM-PAW (excluding $R_a^{BM}$);
    \item[$\bullet$] $R_a^{PAW}$: The attacker’s total reward in PAW;
    \item[$\bullet$] $R_a$:	The attacker’s extra reward in BM-PAW compared with PAW;
    \item[$\bullet$] $R_t^{BM-PAW}$: The target pool’s total reward in BM-PAW;
    \item[$\bullet$] $R_t^{PAW}$: The target pool’s total reward in PAW;
    \item[$\bullet$]$R_t$: The target pool’s extra reward in BM-PAW compared with PAW.
\end{itemize}

\subsection{}\label{apx3}
We define the following parameters to analyze the two-pool BM-PAW attack game:\\

\begin{itemize}
    \item[$\bullet$] $\alpha^{[i]}$: The total mining power of Pool$_i$;
    \item[$\bullet$] $f_{1}^{\lbrack i\rbrack}$: The original infiltration mining power of Pool$_i$;
    \item[$\bullet$] $f_{2}^{\lbrack i\rbrack}$: The reallocated infiltration mining power of Pool$_i$;
    \item[$\bullet$] $c_{1}^{\lbrack i\rbrack}$: The probability of the Pool$_i$’s withheld FPoW is selected as the main chain in two-branch cases;
    \item[$\bullet$] $c_{2}^{\lbrack i\rbrack}$: The probability of the Pool$_i$’s withheld FPoW is selected as the main chain in three-branch cases;
    \item[$\bullet$] $R_{a}^{\lbrack i\rbrack}$: The reward of the Pool$_i$.
\end{itemize}

\section{Possible Cases in Two-pool BM-PAW Game}
\label{apx4}

\begin{itemize}
    \item[$\bullet$] \textbf{Case 1.} Pool$_i$’s innocent mining finds an FPoW;
    \item[$\bullet$] \textbf{Case 2.} Pool$_i$’s infiltration mining first finds an FPoW (in Pool$_{\neg i}$) and then his innocent mining finds another FPoW;
    \item[$\bullet$] \textbf{Case 3.} Pool$_{\neg i}$’s infiltration mining first finds an FPoW (in Pool$_i$) and then Pool$_i$’s innocent mining finds another FPoW;
    \item[$\bullet$] \textbf{Case 4.} Pool$_1$’s infiltration mining (in Pool$_2$), Pool$_2$’s infiltration mining (in Pool$_1$) and Pool$_i$’s innocent mining find three FPoWs in order;
    \item[$\bullet$] \textbf{Case 5.} Pool$_2$’s infiltration mining (in Pool$_1$), Pool$_1$’s infiltration mining (in Pool$_2$) and Pool$_i$’s innocent mining find three FPoWs in order;
    \item[$\bullet$] \textbf{Case 6.} Pool$_{\neg i}$’s infiltration mining first finds an FPoW (in Pool$_i$) and then other miners (except Pool$_i$ and Pool$_{\neg i}$) find another FPoW;
    \item[$\bullet$] \textbf{Case 7.} Pool$_1$’s infiltration mining (in Pool$_2$), Pool$_2$’s infiltration mining (in Pool$_1$) and other miners find three FPoWs in order;
    \item[$\bullet$] \textbf{Case 8.} Pool$_2$’s infiltration mining (in Pool$_1$), Pool$_1$’s infiltration mining (in Pool$_2$) and other miners find three FPoWs in order;
\end{itemize}
\end{sloppypar}
\end{document}